\documentclass[]{spie}  

\usepackage{amsmath,amsfonts,amssymb}
\usepackage{graphicx}
\usepackage[colorlinks=true, allcolors=blue]{hyperref}

\title{MeerTRAP in the Era of multi-messenger astrophysics}

\author[a]{Kaustubh Rajwade}
\author[a]{Benjamin Stappers}
\author[b]{Christopher Williams}
\author[c]{Ewan Barr}
\author[a]{Mechiel Christiaan Bezuidenhout}
\author[a]{Manisha Caleb}
\author[a]{Laura Driessen}
\author[a]{Fabian Jankowski}
\author[a]{Mateusz Malenta}
\author[a]{Vincent Morello}
\author[a]{Sotirios Sanidas}
\author[a]{Mayuresh Surnis}
\affil[a]{Jodrell Bank Centre for Astrophysics, University of Manchester, Oxford Road, Manchester M13 9PL, UK}
\affil[b]{Astrophysics, Denys Wilkinson Building, University of Oxford, Keble Road, Oxford OX1 3RH, UK}
\affil[c]{Max-Planck-Institut f\"ur Radioastronomie, Auf dem H\"ugel 69, D-53121 Bonn, Germany}

\authorinfo{Further author information: (Send correspondence to Kaustubh Rajwade)\\Kaustubh Rajwade.: E-mail: kaustubh.rajwade@manchester.ac.uk \\ Christopher Williams.: E-mail: christopher.williams@physics.astro.ox.ac.uk}

\begin{document} 
\maketitle

\begin{abstract}
Real-time detections of transients and rapid multi-wavelength follow-up are at the core of modern multi-messenger astrophysics. MeerTRAP is one such instrument that has been deployed on the MeerKAT radio telescope in South Africa to search for fast radio transients in real-time. This, coupled with the ability to rapidly localize the transient in combination with optical co-pointing by the MeerLICHT telescope gives the instrument the edge in finding and identifying the nature of the transient on short timescales. The commensal nature of the project means that MeerTRAP will keep looking for transients even if the telescope is not being used specifically for that purpose. Here, we present a brief overview of the MeerTRAP project. We describe the overall design, specifications and the software stack required to implement such an undertaking. We conclude with some science highlights that have been enabled by this venture over the last 10 months of operation.
\end{abstract}


\section{INTRODUCTION}
\label{sec:intro}  
Fast Radio Bursts (FRBs) are bright, short duration radio flashes of as yet unknown origin. Since their discovery, more than a hundred have been published though over 500 of them have been discovered (CHIME/FRB, priv. communication). Their extremely high brightness temperatures and extragalactic distances make FRBs excellent probes for studying important events in the Cosmic history of the Universe. In spite of significant progress made in the field in the last decade, several fundamental questions about the origins of FRBs remain unanswered. The high radio luminosity associated with FRBs and the (initial) lack of further bursts from the first FRBs invoked many cataclysmic astrophysical phenomena to explain their genesis. A variety of cataclysmic models like SuperNovae, HyperNovae and Neutron Star (NS) collapse were initially proposed to explain the properties of FRBs~\cite{yamasaki2018,lin2020}. The discovery of repeating FRBs however, showed that the progenitor does not necessarily get destroyed during the creation of the FRBs. The current consensus on the progenitors of FRBs is under contention with multiple models that can explain the repeating as well as the one-off FRBs. Most of them predict that much of the energy released from an FRB is carried away by optical and high energy photons~\cite{lyutikov2016}, rendering the models less constrained based on only a detection at radio wavelengths. Hence, detection of a contemporaneous multi-wavelength counterpart to an FRB are crucial to probe the physics at the heart of the FRB emission mechanism. 

One of the most promising models to explain one-off FRBs is a NS-NS merger~\cite{lin2020}, which are now known to be excellent sources of Gravitational waves (GWs)~\cite{evans2017}. A number of authors have suggested models where the FRB is created simultaneously with the emission of GWs. The possibility of finding FRBs simultaneously with a GW holds exciting prospects for current and future observing facilities. Multi-messenger astronomy came to the fore a few years ago with the emergence of LIGO and VIRGO GW detectors~\cite{ligovirgo2015}. The first detection of a GW from a neutron star merger combined with the contemporaneous detection of electromagnetic radiation~\cite{abbott2017,evans2017,hallinan2017} showed us that it is possible to study new types of astrophysical transients by exploiting the synergy between multi-messenger and multi-wavelength transient detection facilities. 

MeerTRAP is a European Research Council (ERC) funded project deployed on the MeerKAT radio telescope in South Africa. The instrument has been built with the main aim of finding and localizing transients and FRBs in real-time using the excellent sensitivity of MeerKAT. An added advantage of this project is the possibility of simultaneous optical coverage with the MeerLICHT telescope~\cite{bloemen2016}. This places MeerTRAP in an ideal position to study fast radio transients while looking for optical counterparts and potential radio and/or optical afterglows that will be crucial to answer fundamental questions about FRBs and their relationship to other multi-messenger transients. MeerTRAP will be able to detect and localize any putative FRB from a NS merger, thus confirming a connection between FRBs and GW sources. Below, we present a detailed overview of the MeerTRAP project. The software for the overall system is described in Section~\ref{sec:sof}. The hardware and compute cluster is described in Section~\ref{sec:hard}. The real-time data analysis pipeline is presented in Section~\ref{sec:rtp}. In Section~\ref{sec:comm}, we present the commissioning results of the instrument at the MeerKAT telescope, demonstrating its scientific capabilities.

\section{MeerTRAP Overview}
\label{sec:overview}

\subsection{Hardware}
\label{sec:hard}

 \begin{figure} [ht]
   \begin{center}
   \includegraphics[height=5cm]{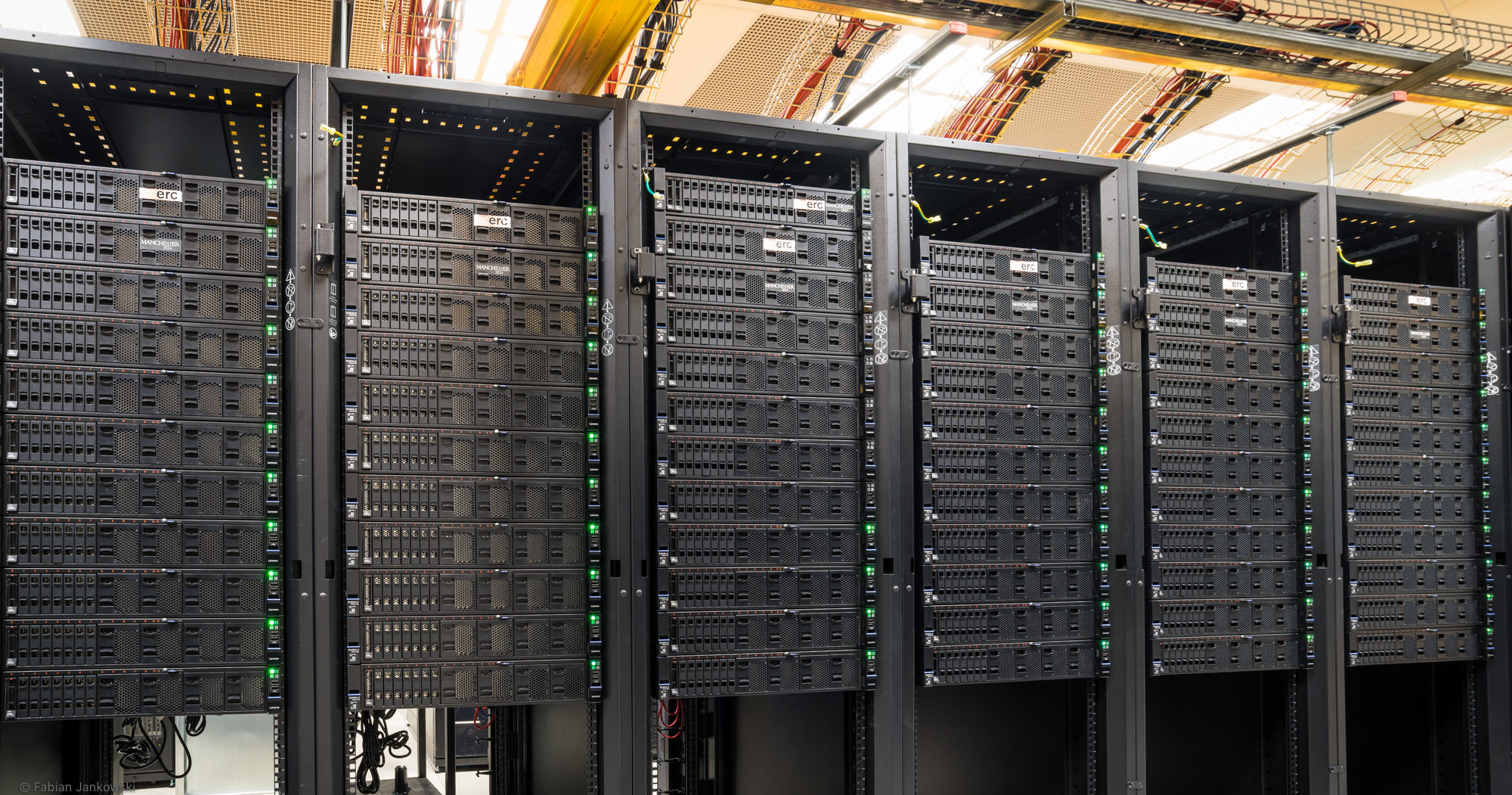}
   \end{center}
   \caption[] 
   { \label{fig:cluster} 
   Picture of the entire MeerTRAP computer cluster installed in the Karoo Array Processing Building.}
   \end{figure} 

Small radio dishes provide a large field of view while larger dishes improve sensitivity and angular resolution at the cost of decreasing field of view. An interferometer can address these limitations by providing the field of view of a small dish and high angular resolution from an array of small dishes spread out over large distances. Sampling of the entire primary field of view of a radio interferometer like MeerKAT (1.27 sq.deg) with highly sensitive coherent beams requires creation of large number of coherent beams. Hence, it is important to have the compute capacity to process every coherent beam in real-time to look for bright radio single pulses. In order to achieve that, heterogeneous compute servers, combining the computational power of multi-core processors and Graphical Processor Units (GPUs) are required. The MeerTRAP compute cluster consists of 66 compute servers with one head node that manages the other 65 compute nodes. Each compute node comprises two Intel Xeon 8C/16T processors each having 16 logical cores at their disposal. They are equipped with two Nvidia GTX 1080 Ti GPUs with 12 GB of global memory in each GPU. Each compute node has a total of 256 GB of RAM and is connected to the network via 10-GbE network interface cards (NICs). All nodes communicate with the head node via 1-GbE ethernet links. The 10-GbE links from each node are connected to a network switch that interacts with the main data transmission network of the telescope via 40-GbE connections. The cluster was procured in early 2018 and then assembled on site in the summer of the same year. Figure~\ref{fig:cluster} shows an image of fully assembled MeerTRAP cluster (also known as Trasient User Supplied Equipment (TUSE)). 

\subsection{Software}
\label{sec:sof}
Due to the compute requirements and commensal nature of MeerTRAP, a complex and robust control and processing software suite had to be developed that was well tested and fully automated. This involved the development of software in multiple programming languages (Python, C++ and CUDA) that were better suited for different functions of the project.
The high-level control system of the MeerTRAP processing cluster is written in Python and acts as the interface between the central telescope management authority of MeerKAT (called CAM, for Control and Monitoring) and the lower-level data ingestion and processing pipelines on the MeerTRAP compute nodes. The exchange of information between CAM and all MeerKAT telescope components, including the MeerTRAP cluster, is based on the \textsc{katcp}\footnote{\url{https://github.com/ska-sa/katcp-python}} protocol which consists of human-readable text messages with a standardized syntax. The MeerTRAP head node is managed by a server dubbed the MasterController that listens for katcp requests sent by CAM, which signal the progress through various stages of an observation. The MasterController is responsible for actively retrieving all relevant configuration parameters from CAM (number of coherent beams, number of frequency channels and time sampling interval of the data, etc.) at the beginning of an observation, and dispatching the processing load between compute nodes optimally; it is also in charge of forwarding real-time metadata changes received from CAM to the compute nodes, including in particular the positions of the coherent beams on sky. Every compute node is managed by its own \textsc{katcp} server called NodeController, whose task is to configure, start and stop the processing pipeline whenever necessary, for example when the subset of beams assigned to the node move to new targets. This involves translating metadata received from CAM via the MasterController into configuration files for every executable program in the pipeline software stack. The NodeController actively tracks the pipeline's progress and informs external observers of its status in real time by exposing a set of so-called \textsc{katcp} sensors, which are key-value pairs that can be read remotely by a \textsc{katcp} client. The MeerTRAP control system also includes a web server that monitors all compute node sensors and displays thre resulting data in real time on dynamic web pages, including a summary table of basic node health indicators, current shape of the bandpass, and thumbnail images of the transient candidates recently produced.

The main data ingest and processing pipeline is built using the C++11 highly templated and highly typed software suite called \textsc{Panda}~\footnote{\url{https://gitlab.com/SKA-TDT/panda}} and \textsc{Cheetah}~\footnote{\url{https://gitlab.com/SKA-TDT/cheetah}}.
\textsc{Cheetah} is an open source C++11 project focusing on providing quasi-real time pulsar and transient processing. Its main design goal has been to provide a platform for prototyping algorithms in such a way that enables the ideas of continuous integration and deployment on heterogeneous systems. Thus a cheetah pipeline should be deployable directly as a real time telescope backend, as well as on a laptop for local data analysis and prototyping without any code modification. A pipeline in \textsc{Cheetah} is composed of a series of functional modules each with well defined interfaces. Each module may have any number of algorithms, perhaps tailored to specific accelerators such as an FPGA or GPU, that can be used to achieve the required functionality. Ideally each module will have a set of generic functional tests that any such algorithm would be expected to pass. The choice of which algorithm to use can be a runtime decision via a command line switch or the configuration file. The cheetah scheduler will then use the most suitable algorithm for the resources that are available at the time, acting as an effective load balancer.

\textsc{Cheetah} relies on the \textsc{Panda} library which which contains a suite of utilities to build generic (i.e. non pulsar/FRB search specific) pipelines. \textsc{Panda} covers numerous mundane tasks as configuration file handling, resource pool management and schedulers, device discovery, data switches, as well as more complex tools such as a UDP packet ingest framework. \textsc{Cheetah} and \textsc{Panda} both use C++ meta-programming extensively. This is a deliberate design decision made in order keep overheads to a minimum whilst maintaining modular flexibility. This comes at the cost of complexity, and the code base requires a relatively high skill level in C++ in order to navigate.

\subsection{Real-time search pipeline}
\label{sec:rtp}

 \begin{figure} [ht]
   \begin{center}
   \includegraphics[height=8cm]{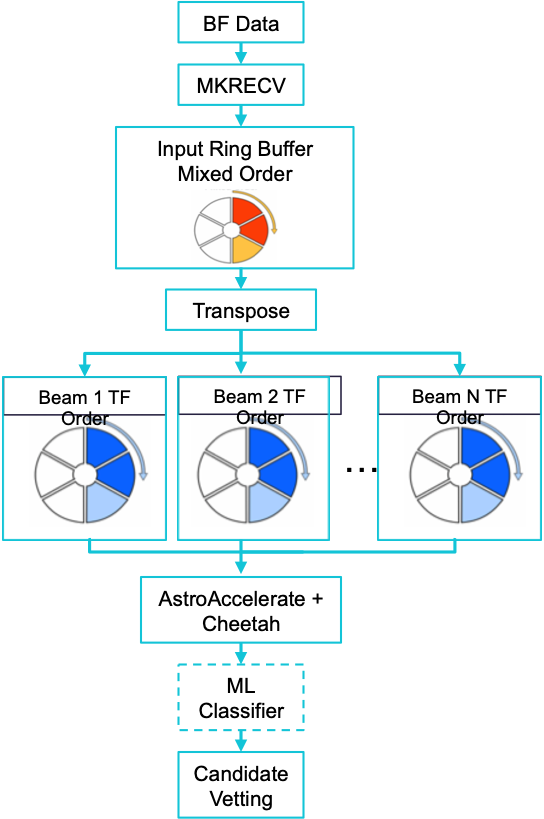}
   \includegraphics[height=8cm]{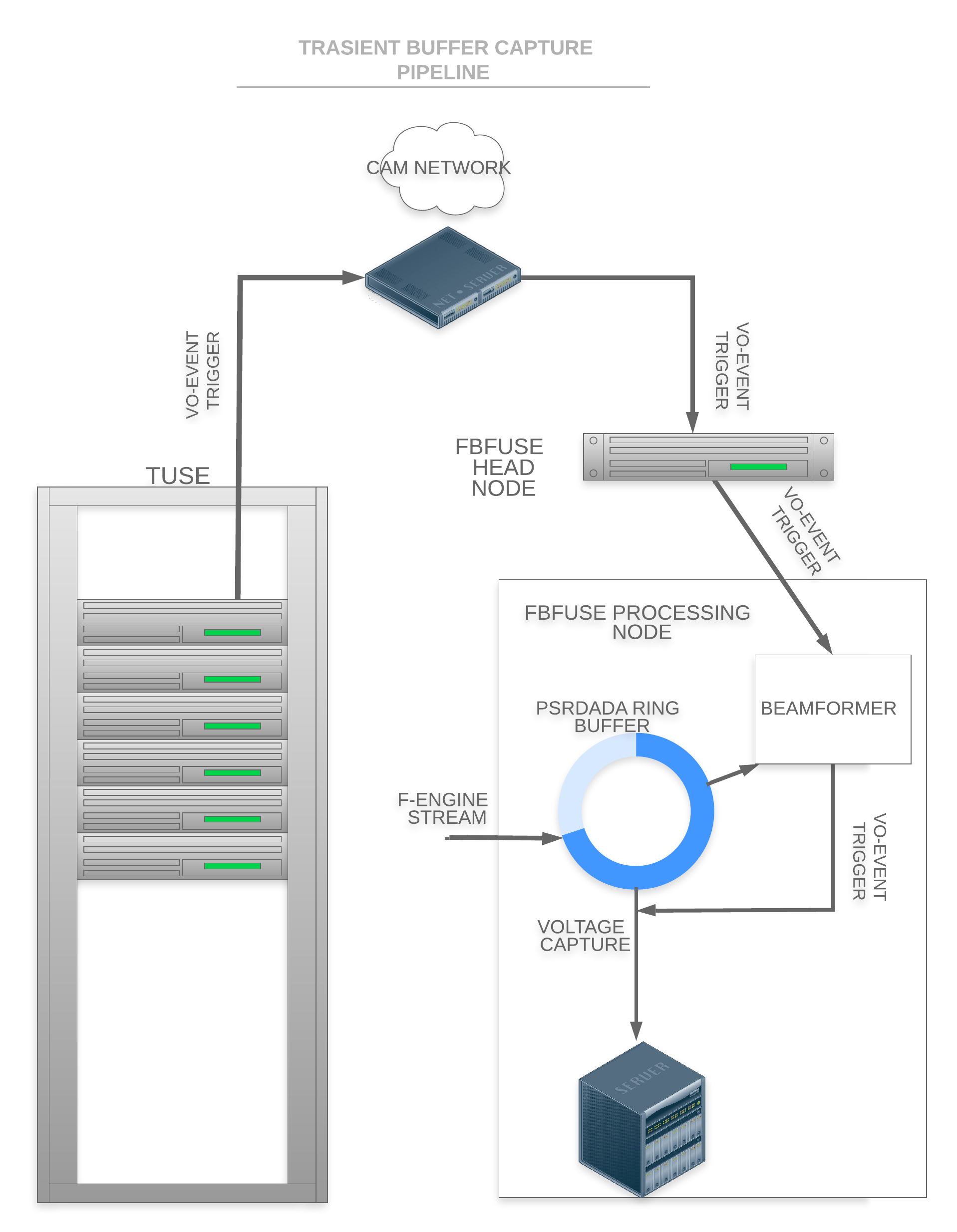}
   \end{center}
   \caption[] 
   { \label{fig:block} 
   \textbf{Left Panel}: Block diagram of the entire real-time single pulse search pipeline of MeerTRAP. \textbf{Right Panel}: Block diagram explaining the transient voltage capture mode of the MeerTRAP system.}
  \end{figure} 

 Figure~\ref{fig:block} shows the block diagram of the full MeerTRAP pipeline. The instrument works in two main modes; 1) Production and 2) Filterbank. In the production mode, data from 32 beamformer nodes of Filterbank BeamFormer User Supplied Equipment (FBFUSE)(a separate compute cluster developed by the Max-Planck-Institut f\"ur Radioastronomie in Bonn, Germany) that produces 768 coherent beams and an incoherent sum of the power from all the available (maximum of 64) antennas, are received via the NIC on each compute node. The data are sent over the network to MeerTRAP via packets that conform to the SPEAD2 protocol~\footnote{\url{https://spead2.readthedocs.io/en/latest/}}. These data streams are then read in by each node using the \textsc{mkrecv}~\footnote{\url{https://gitlab.mpifr-bonn.mpg.de/mhein/mkrecv}} that reads the heaps from the SPEAD2 packets, correctly orders them and saves them to POSIX shared memory ring buffers. Each node reads data corresponding to up to 12 beams in time-frequency format. The data for all 12 beams are then transposed on a per beam basis and the resulting frequency-time format data are saved into separate shared memory ring buffers. From there, the data are read in by the \textsc{Cheetah} pipeline and sent to \textsc{Astro-Accelerate}, a real-time GPU-based single pulse search software~\cite{DimoudiWesley, Karel2, Karel1}~\footnote{\url{https://github.com/AstroAccelerateOrg/astro-accelerate}}. Astrophysical signals are altered by the free electrons along the line-of-sight that makes signals at lower frequencies arrive later than the ones at higher frequencies. This delay is quantified by the dispersion measure (DM) that tracks the integrated electron density along the line of sight. This delay needs to be corrected for before any search for single pulses can be done. The real-time search in MeerTRAP is done by incoherently de-dispersing in the DM range 0--5118.4~pc cm$^{-3}$, divided into multiple sub-ranges with varying DM steps and time averaging factors. We also search up to a maximum boxcar width of 0.67 s in steps of cosecutive powers of 2. A specific choice of parameters allows us to process all the data in real time, thanks to strict optimisations applied in the \textsc{Astro-Accelerate} algorithms. 

To reduce the number of detections due to Radio Frequency Interference (RFI), we applied a static frequency channel mask to the data before the de-dispersion and single-pulse search. More recently, a new RFI mitigation algorithm called Inter-Quartile Range mitigation (IQRM) was implemented and deployed on the pipeline that cleans the data from narrow-band RFI by automatically generating an adaptive, time-variable mask (Morello et al. in prep). That combined with the static channel mask gets rid of most of the corrupted data from our pipeline. Additionally, the data are cleaned using standard zero-DM excision \cite{eatough2009} to remove any remaining RFI that was infrequent enough not to be detected by the IQRM algorithm or is too broad band to be masked by the static channel mask. The extracted candidate files contain raw data of the dispersed pulse and additional padding of 0.5~s at the start and at the end of the file. The resulting candidates are then sifted through a coincidence filter (when the same candidate appears in multiple beams, that is very likely to be RFI)
 and run through a known source matching routine that matches the candidates to known radio emitting neutron stars (pulsars) or other transients in the field. The remaining candidates are vetted visually for astrophysical origin. Currently, a new machine learning based classifier is being implemented in order to classify new astrophysical candidates automatically in real-time.
 
In the Filterbank mode, the beamformed data are written to POSIX shared memory buffers. The data are then transposed on a per beam basis and saved to a file instead of another ingest buffer for the single pulse search pipeline. The file conforms to the standard \textsc{SIGPROC} format for saving dynamic spectrum data~\footnote{\url{http://sigproc.sourceforge.net/sigproc.pdf}}.

Lastly, we have recently deployed a transient buffer capture mode whereby a portion of complex voltage data around the FRB are captured on the FBFUSE nodes in order to study the burst at higher time resolution and to image and localize it to arc-second precision. This mode does depend on the MeerTRAP cluster sending an event trigger to FBFUSE to perform voltage capture in real-time. The voltage extraction algorithm takes care of the dispersion delay due to the large dispersion measure of the FRB and only saves data around the time of the FRB at every frequency. Right panel of figure~\ref{fig:block} shows the schematic of the transient buffer capture code. The details of these pipelines will published in a separate paper (Stappers et al. in prep).

\subsection{Commissioning/ Results}
\label{sec:comm}
\begin{figure} [ht]
   \begin{center}
   \includegraphics[height=8cm, angle=-90]{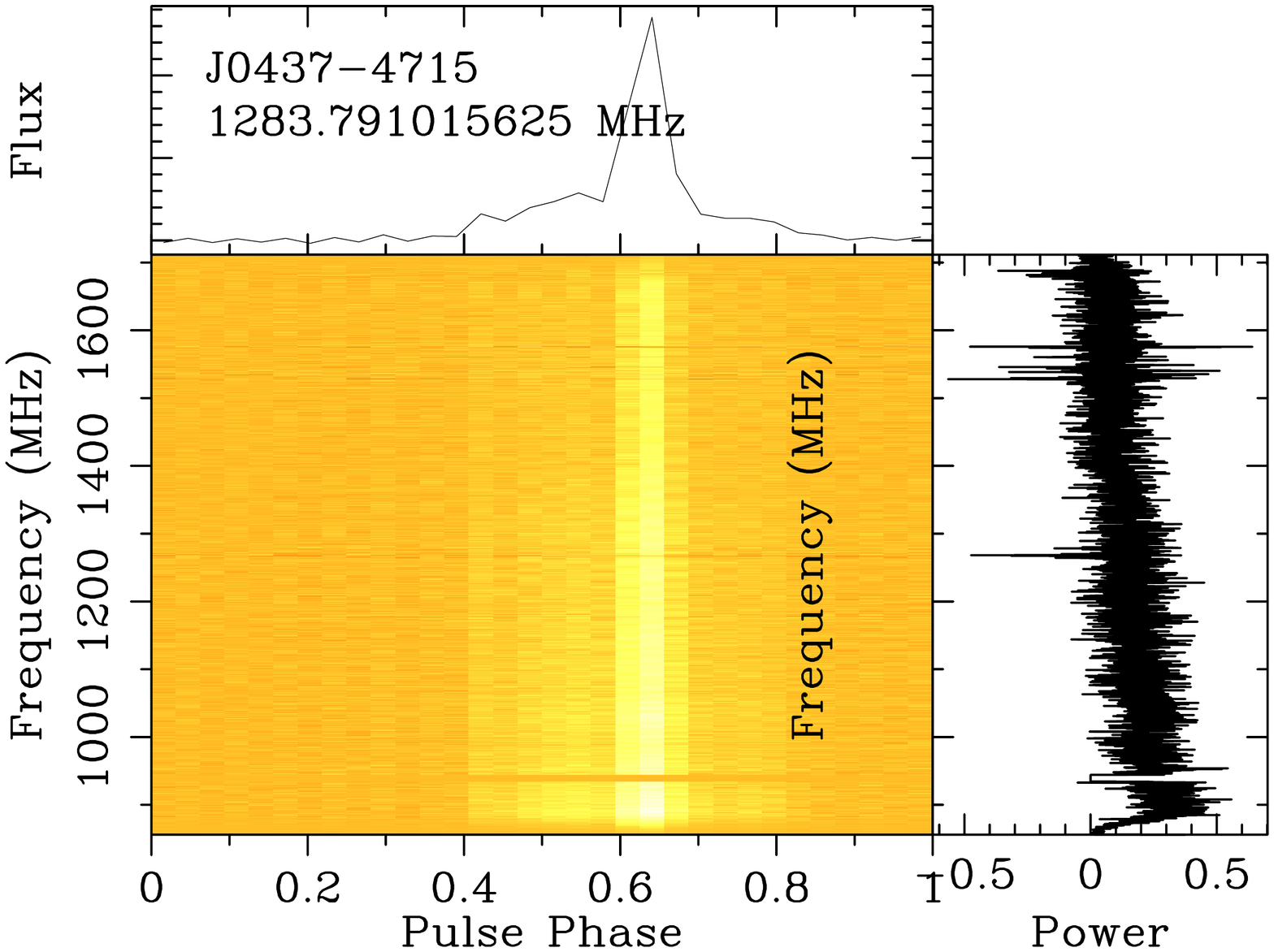}
   \includegraphics[height=8cm, angle=-90]{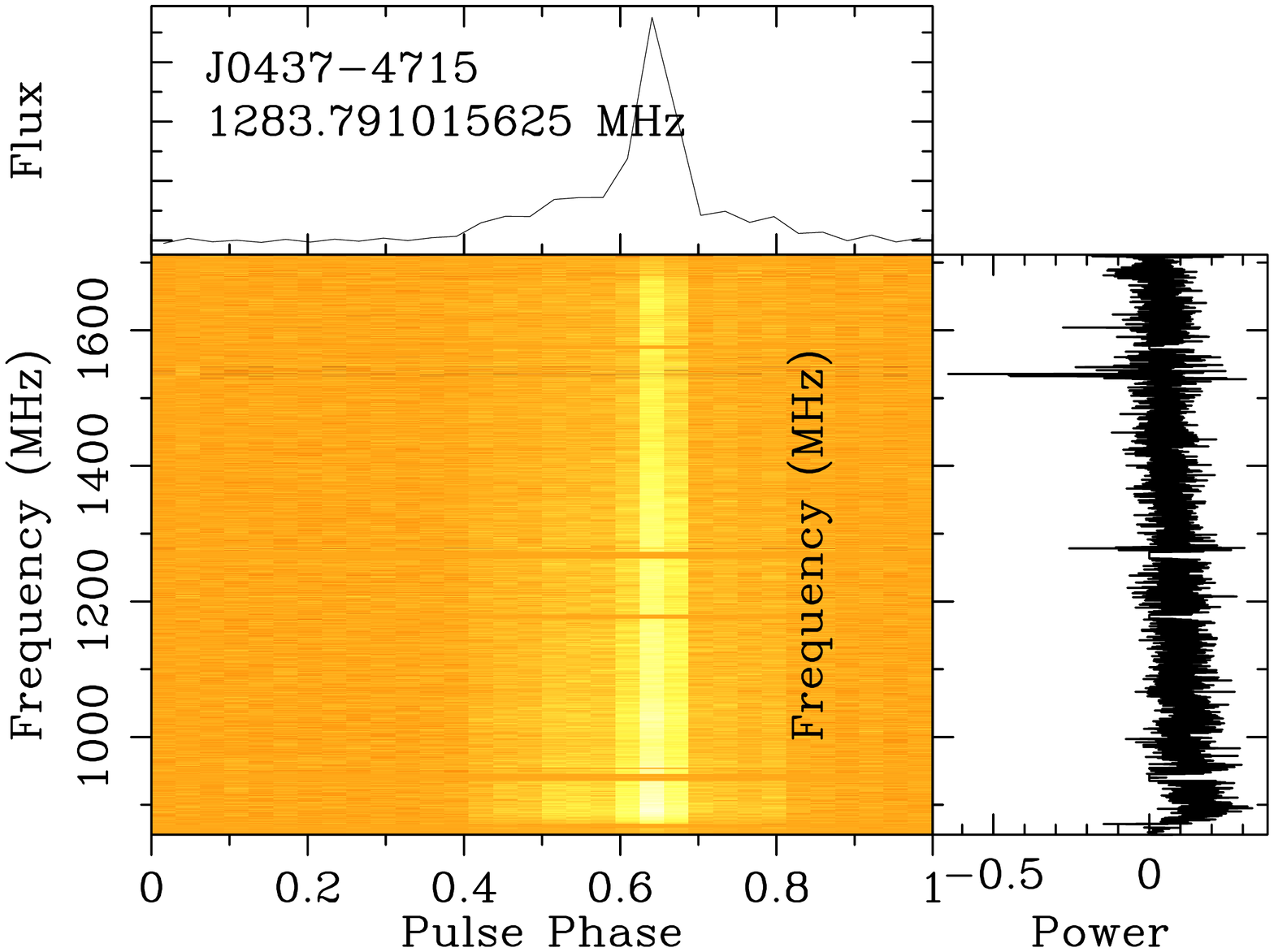}
   \end{center}
   \caption[] 
   { \label{fig:psr} 
   \textbf {Left Panel}: Dynamic spectrum along with the folded pulse profile of PSR J0437-4715 from a coherent beam observation during the commissioning of the MeerTRAP instrument (S/N 488). \textbf{Right Panel}: Same plot from the incoherent beam observation at the same time (S/N 422). Note that the coherent beam was offset from the location of the pulsar by 20 arc-minutes.}
  \end{figure}

\begin{figure} [ht]
   \begin{center}
   \includegraphics[height=8cm]{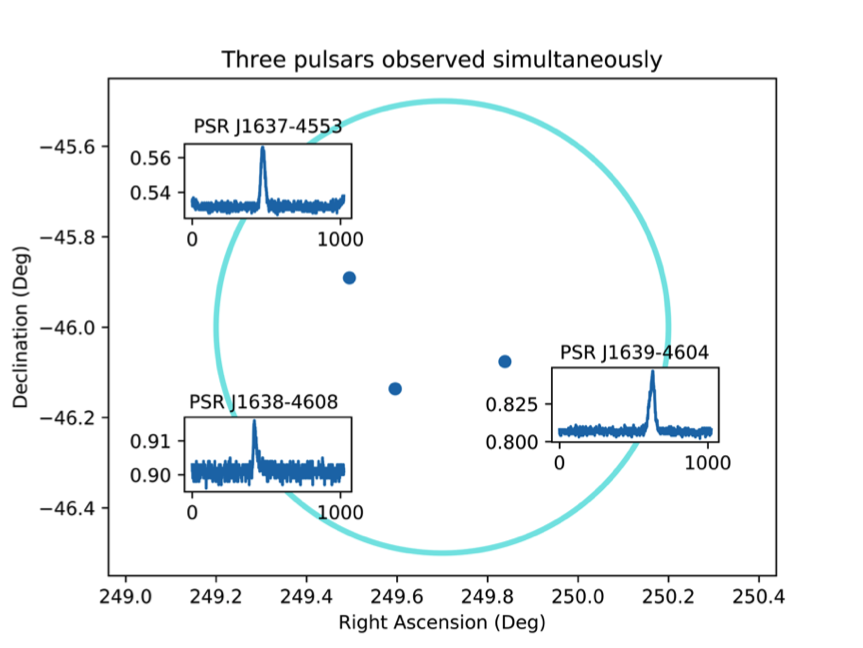}
   \end{center}
   \caption[] 
   { \label{fig:tile} 
Three pulsars detected by observing them simultaneously with MeerTRAP using three separate beam tilings formed by FBFUSE (blue dots) within the primary Field of View (cyan circle). The inset shows the folded profile of the pulse from the data taken by the MeerTRAP cluster.}   
  \end{figure}

The commissioning of the instrument started in April of 2019 where various modes of the system were thoroughly tested. A number of bright know pulsars were observed during these tests. Figure~\ref{fig:psr} shows the folded profile of pulsar PSR J0437-4715 from the recorded data that showed that coherent and incoherent beam were created correctly and the pulsar was detected with the expected significance in both the beams. We were able to show that the system was more than capable of detecting bright radio pulsars. Another attribute of the MeerTRAP system that was successfully tested was the ability to perform a real-time search in multiple coherent beam tilings within the primary field of view of MeerKAT. Figure~\ref{fig:tile} shows an image where three separate radio pulsars located within the primary beam were observed. Three tilings were successfully made and all three pulsars were detected by MeerTRAP. By doing this experiment, we were able to demonstrate that FBFUSE can form flexible beam tilings upon request from MeerTRAP and the MeerTRAP instrument is able to successfully detect these astrophysical sources. This ability is important in order to follow-up interesting sources and transients when they end up in the MeerKAT FoV during observations.

\begin{figure} [ht]
   \begin{center}
   \includegraphics[height=8cm]{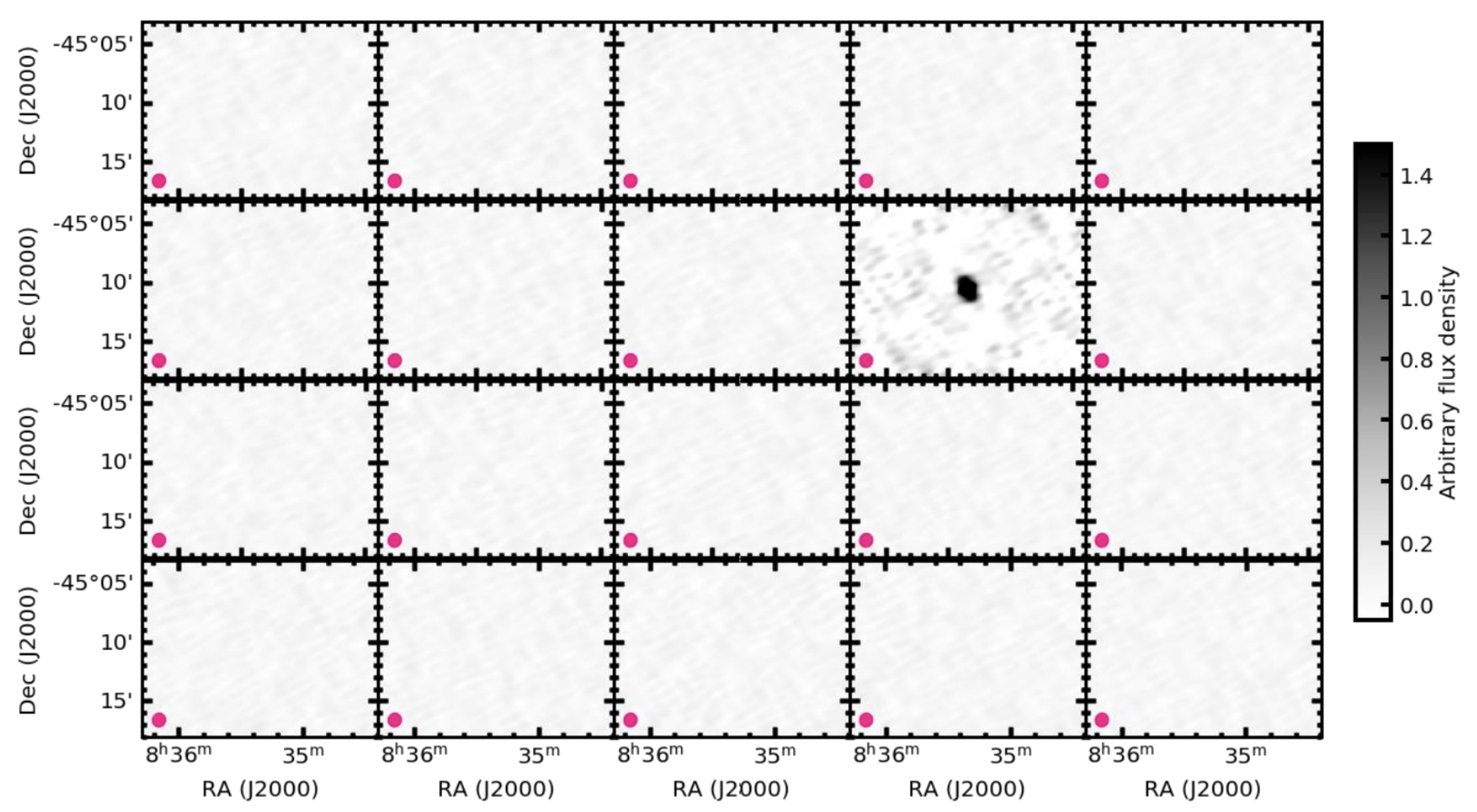}
   \end{center}
   \caption[] 
   { \label{fig:vela} 
Plot shows 20 panels of a radio image made from 100~ms of raw complex voltage data from MeerKAT. Each panel represents 5~ms of integrated data. The Vela single pulse is clearly detected in one panel as the pulse width is less than 5~ms. The pink dot represents the size of the synthesized beam in this image.}   
  \end{figure}

Another important milestone that was achieved recently was the ability to trigger and save complex voltage data on a single radio pulse from a bright pulsar. This mode of operation is crucial in order to localize and study the spectro-temporal structure of FRBs in great detail. Right panel of figure~\ref{fig:block} shows the schematic diagram of the transient voltage capture mode. In the event of a confirmed detection of an FRB by MeerTRAP, an event trigger consisting of all the relevant information about the new FRB will be sent to FBFUSE that will enable a capture of complex channelized voltage data on the filesystem. These data are then analysed offline and imaged. To test this system, we observed the Vela pulsar (PSR B0835-45) and then and a manual trigger was sent to FBFUSE in order to activate a voltage capture of the data. These data were analysed offline to make an image and a coherent timeseries. Figure~\ref{fig:vela} shows an image of the Vela single pulse data generated from the channelized complex voltages that were saved from the trigger that shows a clear detection of the pulse. This exercise showed us that we can trigger and capture voltage data of a detected transient in order to image the data and localize it.

The first commensal science observations with the real-time system started in August of 2019 where we observed with a pulsar timing project. Since then, the majority of our time so far has been spent in the Galactic Plane. This has resulted in a number of discoveries of Galactic Transients (Bezuidenhout et al. in prep) and new FRBs (Rajwade et al. in prep). The system has also been used for targeted follow-up of interesting radio transients like FRB 121102~\cite{caleb2020} and the galactic magnetar SGR J1935+2154 (Stappers et al., in prep.).

MeerTRAP has an additional advantage in terms of obtaining multi-wavelength coverage to search for transients. There are opportunities to get simultaneous optical coverage with the 0.65~m MeerLICHT telescope in Sutherland, South Africa. MeerLICHT can provide a 2.7 sq.deg of simultaneous optical sky coverage that covers the entire the FoV of MeerKAT. It does 60-second snapshot of the sky and can go as deep as 20$^{\rm th}$ magnitude in the SDSS-g' band. This holds exciting prospects for MeerTRAP as there is potential to find a prompt optical flash in the direction of an FRB detected by MeerTRAP or an optical afterglow, providing tight constraints on the progenitors of FRBs and probing the connection between FRBs and GW sources. Robotic operations of synchronized MeerLICHT and MeerKAT observations have recently begun.

\section{Summary}
Here, we present a detailed overview of the MeerTRAP instrument. The real-time FRB detection system comprises of an heterogeneous code base with all the major components programmed in templated C++11 and the control system is fully deployed using the standard Python communication libraries. The system is operational at the MeerKAT telescope. Commensal science operations have started since late 2019 and the instrument has already started to find new FRBs and Galactic Transients. As MeerKAT becomes commensal with more and more Large Survey Projects that have been allocated time at MeerKAT, it will detect more and more interesting transients. The deployment of MeerTRAP and the simultaneous optical coverage with MeerLICHT is extremely timely in the era of Multi-Messenger Astronomy and it will be able to deliver high-impact scientific results that probe FRB astrophysics and their relationship with GW sources. 

\subsection{Acknowledgments}
This project has received funding from the European Research Council (ERC) under the European Union’s Horizon 2020 research and innovation programme (grant agreement No. 694745).

\bibliography{report} 
\bibliographystyle{spiebib} 

\end{document}